# Study of neutron response and n-γ discrimination by charge comparison method for small liquid scintillation detector

J. Cerny[a], Z.Dolezal[a], M.P.Ivanov[a*], E.S.Kuzmin[b], J.Svejda[a], I.Wilhelm[a]

[a]*Institute of Particle and Nuclear Physics, Faculty of Mathematics and Physics, Charles University, V Holesovickach 2, CZ-18000, Prague 8, Czech Republic*
[b]*Joint Institute for Nuclear Research, Laboratory of Nuclear Problems, Dubna, Moscow Region, 141980 Russia*

--------------------------------------------------------------------------------

**Abstract**: The study of the neutron response and n-γ discrimination for small (18 x 26 x 8) $mm^3$ liquid scintillator BC501A (Bicron) detector was carried out by digital charge comparison method. Three ranges of neutron energies were used: uniform distribution from 0.95 MeV to 1.23 MeV, continuous spectra of AmBe source and monoenergetic 16.2 MeV neutrons. The obtained results are compared with those for cylindrical liquid scintillation detector (40 mm diameter, 60 mm length) at the same energies of neutrons. A dramatic fall of the neutron response function at 400 keVee for small detector at 16.2 MeV neutron energy was measured. For (0.95 – 1.23) MeV neutron energy range such fall takes place at 260 keVee. The greater slope of neutron locus at (0.95 – 1.23) MeV neutron energy comparing to 16.2 MeV for both detectors is explained by longer tail of pulse from proton recoils within (0.1-1.23) MeV energy range.

PACS: 29.90. + r.28.20.Cz

Keywords: Neutrons; Neutron detection;

--------------------------------------------------------------------------------

*Corresponding author:
phone: +420 22191 2768; fax: +420 22191 2434
e-mail: minati@mail.ru

## 1. Introduction

In the experimental research of 3N interactions with polarized neutron beam and polarized deuteron target [1,2,3] at an energy range 14-17 MeV the use of liquid scintillator neutron detectors is proposed for monitoring of neutron beam. In previous setup [4] plastic scintillators were applied as monitor detectors. The use of BC-501 A (BICRON) liquid scintillator gives an efficient way to discriminate between incident gamma rays and neutrons by means of pulse shape discrimination (PSD). We apply a PSD method, which is based on charge integration of the pulse current over two different time intervals using charge-integrating QDC. To minimize the original neutron spectrum distortion for the experimental facility the size of monitor detector has to be as small as possible. In this paper the study of prototype thin monitor detector with inner dimensions (18 x 26 x 8) $mm^3$ is described. As a reference detector cylindrical liquid scintillator volume with dimensions 40 mm diameter, 60 mm length was used. The measurements were made for three ranges of neutron energies: (0.95 - 1.23) MeV, continuous spectra of AmBe source and monoenergetic 16.2 MeV neutrons. Method to obtain experimental response functions and to estimate n-γ discrimination is described.

There are a lot of fields (for example in fusion experiments), where it has become common to use small volume detectors to minimize the neutron spectrum distortion in the experimental facility [5].

## 2. Experimental method.

### 2.1. Scintillation counters.

The study was carried out with the prototype of monitor scintillation counter with 18 mm width, 26 mm length, 8 mm thickness intrinsic volume of BC501A (Bicron) liquid scintillator encapsulated in aluminums cell with 1.5 mm thickness walls with polished surface. The scheme of counter is shown on fig 1. The scintillation cell was separated with 8 mm thick quartz window from 20 cm long Perspex light guide (4 cm of diameter) connected to a Philips XP-2020 (5 cm diameter) photomultiplier tube. The light guide was used to avoid the undesired influence of the magnetic field on photomultiplier tube and to obtain enough room for setup of two monitor counters in future experiments on the way of neutron beam. For this purpose scintillation volume was made 5 mm apart of axes of counter. For light reflection light guides was covered with thin film of aluminized Mylar.

The second scintillation counter consisting of 4 cm in diameter and 6 cm in depth BC501A liquid scintillator encapsulated in aluminum cell with white reflecting paint applied inside was used as a reference detector. This detector was also coupled to the Philips XP-2020 photomultiplier tube.

### 2.2. Experimental setup.

The neutrons of the energy of (0.95 – 1.23) MeV were produced as secondary particles from $^3H(p,n)^3He$ reaction on the Van de Graaf electrostatic accelerator HV2500AN. Protons at the energy of 2.0 MeV bombarded Ti-T target (2 mg/cm$^2$) on a molybdenum backing. The target was mounted at an angle 45° to the incident beam of protons, so the real thickness of the target was 2.82 mg/cm$^2$. Due to approximately linear energy loss of protons in the target, neutron energies had a uniform distribution from 0.95

MeV up to 1.23 MeV. The neutron detectors were mounted at a distance of 3 m from the tritium target and at an angle of $5^O$ from the proton beam axis.

Neutrons with continuous spectra from about 0.5 MeV up to about 10 MeV were obtained with AmBe neutron source. Intensity of the source was of $10^6$ neutrons /sec. To lower intensity of γ-rays with energies 40-60 keV accompanied α-particle emission of americium a lead plate with 5 mm thickness was installed in front of scintillation counters.

For production of neutrons at the energy of 16.2 MeV $^3H(d,n)^4He$ reaction was used. A deuteron beam with $E_d$ = 1.82 MeV was striking Ti-T target (2mg/cm$^2$) at the same setup. To achieve a monoenergetic collimated neutron beam the associated particle method was used [4]. The recoil alpha- particles emitted at the laboratory angle of $90^O$ with respect to the primary deuteron beam were registered with semiconductor surface-barrier (SSB) detector. The neutrons with an energy of (16.2 ± 0.1) MeV associated with these alpha- particles were emitted at an angle of $(62.0 ± 0.7)^O$. The neutron detectors were placed at that angle at the distance of 38 cm from the target. Only neutrons in coincidence with the recoil alpha- particles were registered. The collimators inside the target chamber for the recoil alpha- particles and the mask on the SSB detector face have provided with collimated beam of neutrons with FWHM about 10 mm on neutron detectors .

2.3. Electronics

Fig.2 presents the block diagram of the electronics used in the study. The anode signal is fanned out into three identical signals, by using Linear FAN IN/OUT. Two of these signals are given different delays and then sent to their respective QDC channel inputs. The third signal is delayed and passed through constant fraction discriminator (CFD) to produce a logical flag to indicate that an event has occurred and to trigger the gate signal to the QDC.
The typical time relations between the photomultiplier pulse and gates at the inputs of QD converter are shown in the lower part of fig.1. An initial delay of 16 ns and width of gates 150 ns were found to be optimal for the n-γ discrimination for both detectors. The dotted line in fig.2a shows the scheme, when the experiments with accelerated protons and AmBe source were performed and the associated particle method was not used.

3.Results and discussions

Plotting the charge in the tail of pulse against the total charge leads to the two-dimensional spectra in Figs.3(a, b, c), 4(a, b, c) for small and reference detectors respectively. The Figs.5 (a, b) are expansions of the first few channels of Figs.3(a) and 4(a). One can see the separation between neutron (upper) and gamma-rays (lower) events in the figures. In order to see the separation between neutron and gamma-rays more clearly one-dimensional spectra were produced by projecting onto the ordinate the data in the slices in figs.5(a, b). Owing to these spectra we can determine the lower point of separation between neutrons and gamma-rays ranges, and making contours around these ranges as in Fig. 4(a) to obtain one dimensional spectra of neutron locus that is to determine experimental neutron response function (Figs.3,4(d,e,f)) and to calculate neutron and gamma-rays counts in respective locus. The energies in keVee (keV of electron energy) in Figs.5(c,d) are the result from a calibration of channel number versus electron energy obtained by measuring of known Compton edges of several gamma-ray sources.

The spectra at right parts of Fig.3 give the possibility to compare the neutron response at (0.95 – 1.23) MeV and 16.2 MeV neutron energies and at continuous spectra of energy (0.5 - 10 )MeV from AmBe source for small detector at the same high voltage on photomultiplier. One can see on Fig.3(f) a dramatic fall of the neutron response function to lower level at 400 keVee for 16.2 MeV neutrons. In this case most of recoiled proton have had too large energy and a small stopping power. For neutrons with maximum energy 1.23 MeV this fall takes place at 260 keVee (Fig.3(a)). It is only about twice less comparing to the fall for 16.2 MeV neutrons. At the same time AmBe spectra Fig.3(e) have a continuous long tail without any fall. In that case the recoiled protons have had small energy and stopped in the detector volume. Practically the same one and two dimensional spectra from AmBe source are seen in Figs.3(b, e) for small detector and in Figs.4(b, e) for reference detector.

The scattering process of neutrons on protons at the energy of neutrons up to 10 MeV, is isotropic in the center -of-mass coordinate system. So the response function of a detector based on simple hydrogen scattering of monoenergetic neutrons in this energy range should have a rectangular shape [6]. But due to the relatively large energy spread of neutrons in T + p reaction (0.95 – 1.23) MeV and another distortion mechanisms [7] one can see a significant deviation of one dimensional spectra from pure rectangular shapes in Fig.3 (d) and Fig.4 (d).

The spectra on Figs.4 (a, c) are obtained at the same applied high voltage for reference detector. Comparing the two dimensional spectra at (0.95-1.23) MeV and 16.2 MeV neutron energy on Figs.4(a, c) one can see remarkable difference in the slope of neutron locus. The same picture is observed for small detector – Figs.3(a,c). At the beginning of T+d and AmBe neutron spectra the slope is greater for both detectors. This phenomena can obviously be explained by longer tail of pulse from proton recoils within (0.1-1.23) MeV energy range.

To determine the quality of n-γ discrimination all obtained two-dimensional spectra were separated into three parts – neutron, gamma-ray and common and number of events in every part Nn, Nγ and Nc were calculated (see Fig.4(a)). The ratios Nc/Nn, Nγ/Nn are shown in first and second raw respectively in table 1.
The criteria Nc/Nn allow to take into account a loss of events due to pile up of neutrons and gamma-rays and noise at the beginning of spectra. As one can see in table 1 the loss of events Nc/Nn for small detector is substantially higher for AmBe and T+p neutrons comparing to the T+d neutrons when alpha-particle –neutron coincidences are used for trigger (last two columns). At the same time due to small volume Nγ/Nn is lower for small detector comparing to the reference ratio.

**4.Conclusions**

Pulse shape discrimination method, which is based on charge integration of the pulse current over two different time intervals using charge integrating QDC, was applied to study the neutron response and n-γ discrimination for small detector (18x26x8)mm$^3$ with liquid scintillator BC501A (Bicron). At 16.2 MeV neutron energy a dramatic fall of the neutron response function at 400 keVee was observed. For (0.95-1.23) MeV neutron energy the fall takes place at 260 keVee. The greater slope of neutron locus at (0.95-1.23) MeV neutron energy comparing to 16.2 MeV is explained by longer pulse tail from proton recoils within (0.1-1.23) MeV energy range.

**Acknowledgements**
The authors express their gratitude to J.Broz, Yu.A.Usov, N.S.Borisov, V.G.Kolomiets for their help. The financial support from from the Grant Agency of the Czech Republic under the Grant Nrs.202/00/0899 and 202/03/0831 is gratefully acknowledged.

Table 1.

Comparison of the quality of n-γ discrimination for small and reference detector for different neutron sources. See text for more details.

|  | AmBe Small detector | AmBe Refer. detector | T+p Small detector | T+p Refer. detector | T+d Small detector | T+d Refer. detector |
|---|---|---|---|---|---|---|
| Nc/Nn | 10.2 | 2.21 | 11.7 | 1.45 | 0.85 | 0.75 |
| Ng/Nn | 1.55 | 1.59 | 1.47 | 0.72 | 0.15 | 0.44 |

**Figure captions**

Fig.1. Scheme of a small neutron detector. 1) Detector vessel. 2) Liquid scintillator. 3) Quartz window. 4) Light guide. 5) Photomultiplier tube.

Fig.2. The block scheme of the arrangement used to measure the n-γ discrimination by the digital charge comparison method
(Linear FAN IN/OUT- LeCroy428F, delay – Polon1506, QDC- LeCroy2249SG, gate generator-CAEN93B, coincidence- LeCroy465, CFD-TennelecTC454).
   The typical time relation between the photomultiplier pulse and gates at the inputs of the QD converter is shown in the lower part of the figure.

Fig.3. Left side: (a, b, c) two-dimensional spectra of total charge versus charge in the tail of pulse (channels) for small detector. Right side: (d, e, f) one-dimensional spectra - projections of neutron locus onto x-axes (neutron response functions).

Fig.4. Left side: (a,b,c) two-dimensional spectra of total charge versus charge in the tail of pulse (channels) for reference detector. Right side: (d,e,f) one-dimensional spectra, projections of neutron locus onto x-axis (neutron response function).

Fig.5. Expansions of first few channels of two dimensional spectra of Figs.3a, 4a respectively (a, b). Also shown are the positions of three of energy slices in which data are projected onto the y-axes to create the one-dimensional spectra in right part of the figure (c, d). The energies are result from calibration of channel number versus electron energy obtained by measuring of known Compton edges of several gamma-ray sources.

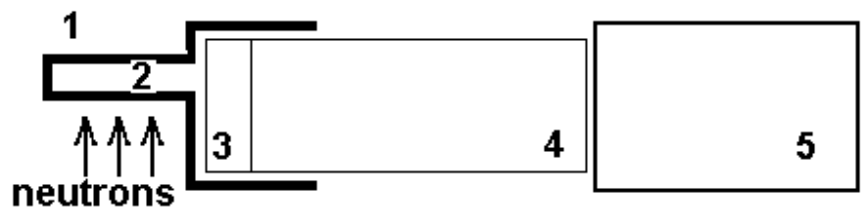

Fig.1

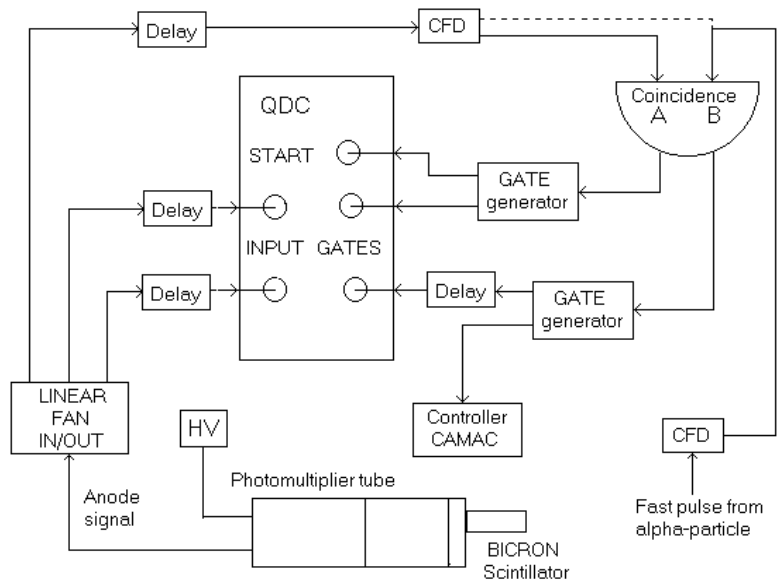

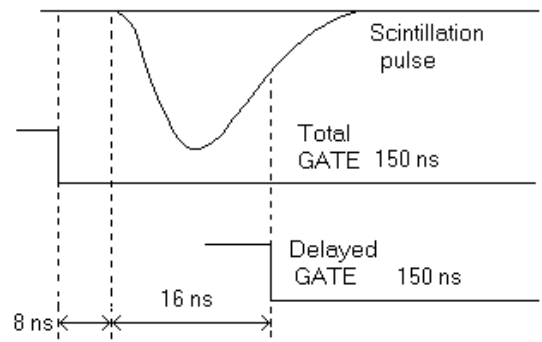

Fig.2.

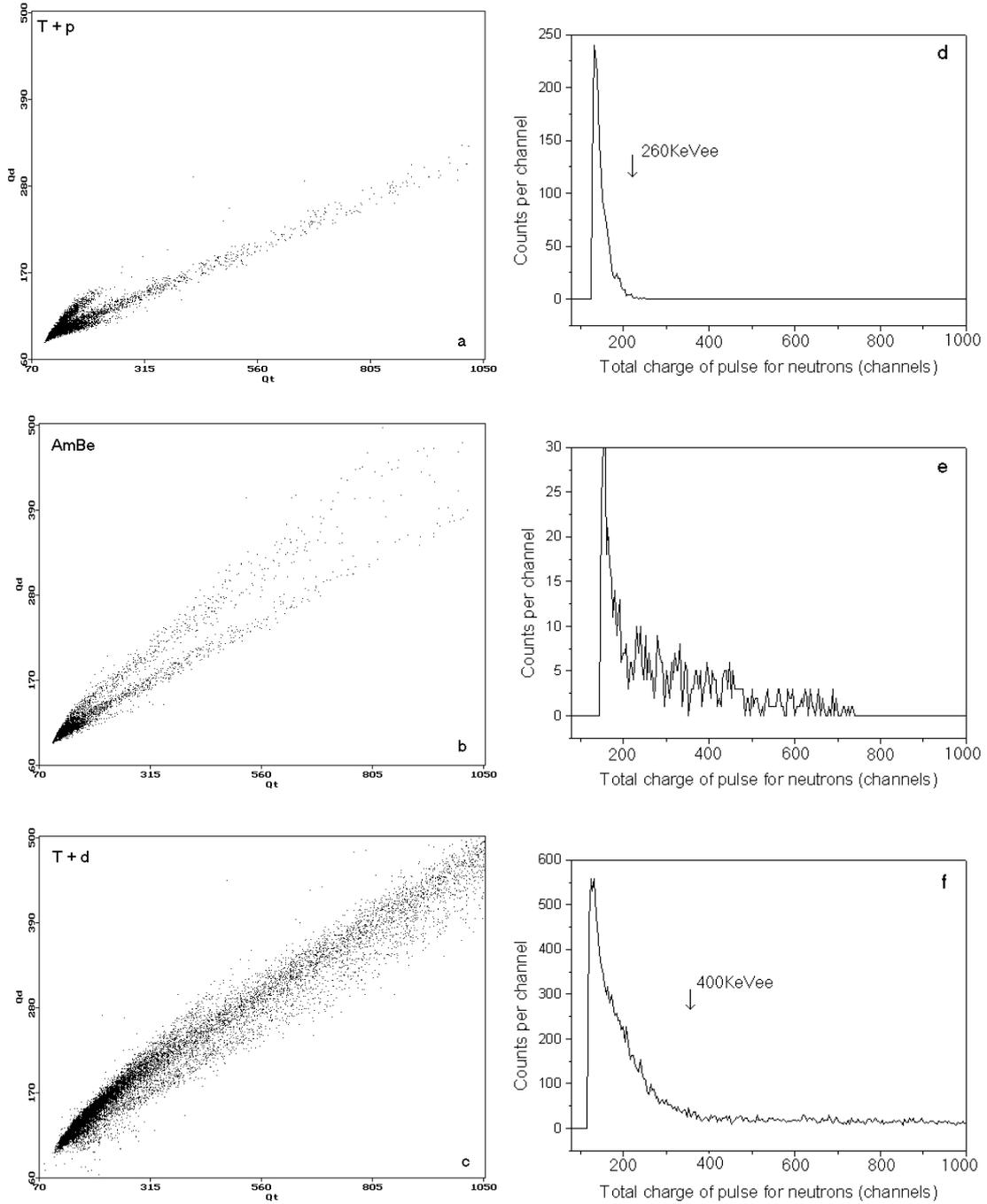

Fig.3.

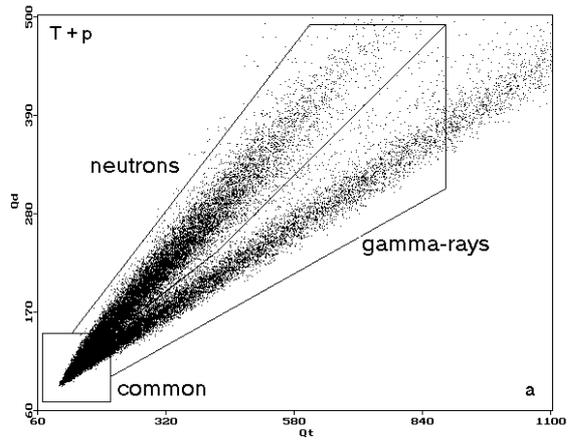
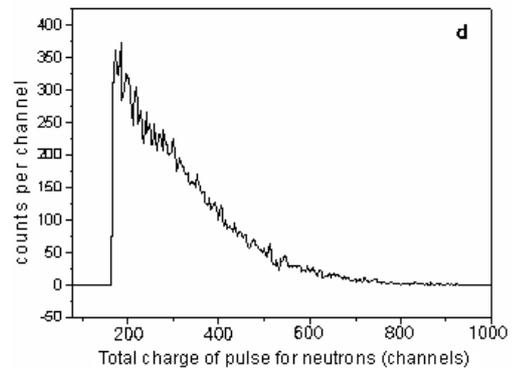
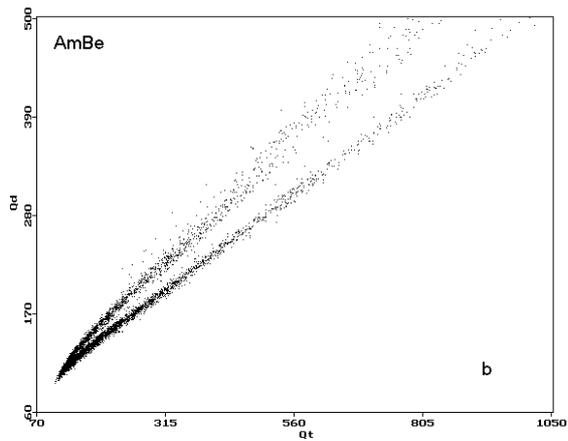
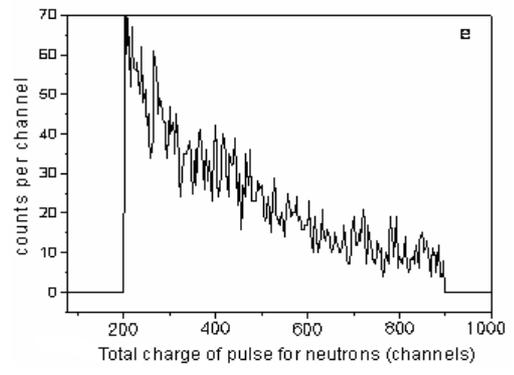
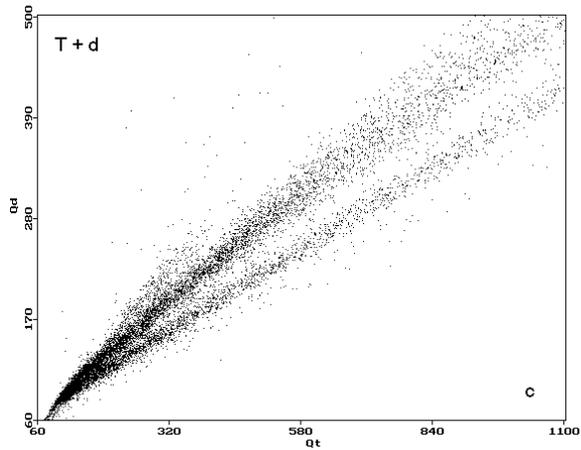
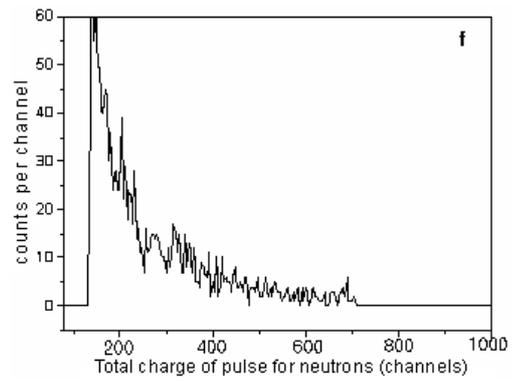

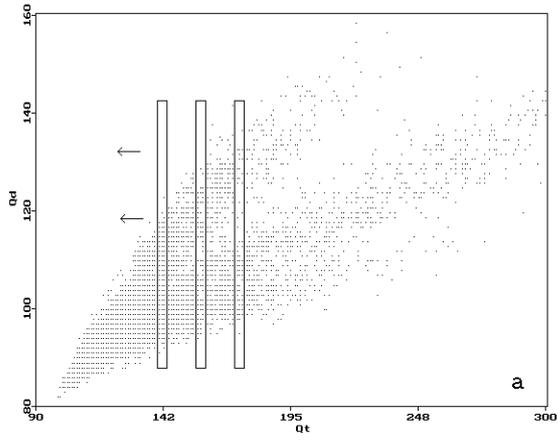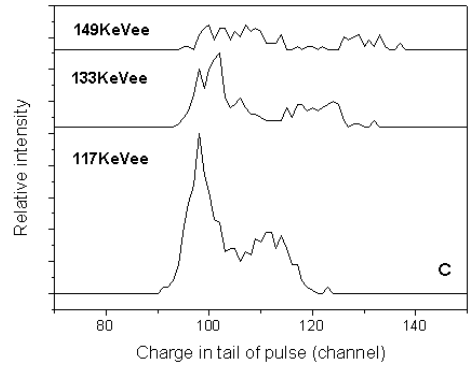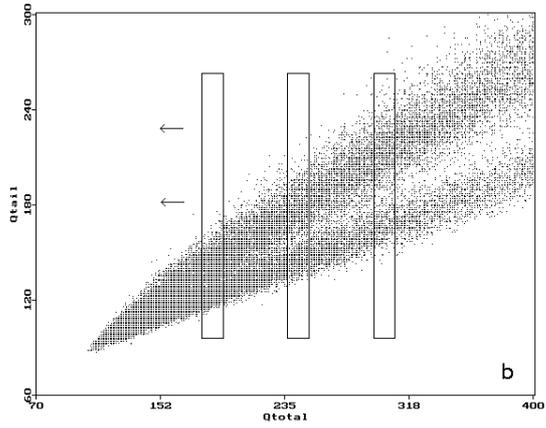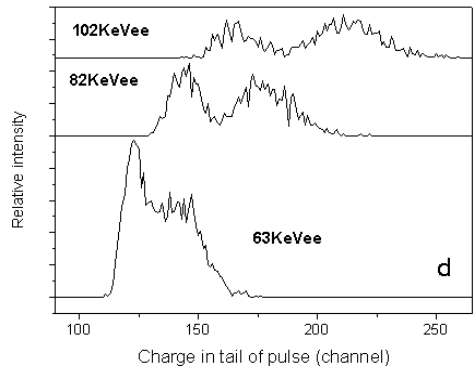

Fig.5.